# Securing HPC using Federated Authentication

Andrew Prout, William Arcand, David Bestor, Bill Bergeron, Chansup Byun, Vijay Gadepally, Michael Houle,
Matthew Hubbell, Michael Jones, Anna Klein, Peter Michaleas, Lauren Milechin, Julie Mullen,
Antonio Rosa, Siddharth Samsi, Charles Yee, Albert Reuther, Jeremy Kepner
MIT Lincoln Laboratory, Lexington, MA, U.S.A.

*Abstract*—Federated authentication can drastically reduce the overhead of basic account maintenance while simultaneously improving overall system security. Integrating with the user's more frequently used account at their primary organization both provides a better experience to the end user and makes account compromise or changes in affiliation more likely to be noticed and acted upon. Additionally, with many organizations transitioning to multi-factor authentication for all account access, the ability to leverage external federated identity management systems provides the benefit of their efforts without the additional overhead of separately implementing a distinct multi-factor authentication process. This paper describes our experiences and the lessons we learned by enabling federated authentication with the U.S. Government PKI and InCommon Federation, scaling it up to the user base of a production HPC system, and the motivations behind those choices. We have received only positive feedback from our users.

*Index Terms*—High Performance Computing, Federated Authentication, Federated Identity Management, Security, Public Key Infrastructure, Multi-Factor Authentication, PKI

## I. INTRODUCTION

The MIT Lincoln Laboratory Supercomputing Center (LLSC) provides a high-performance computing platform to over 1000 users at MIT across several systems and is heavily focused on highly iterative interactive supercomputing and rapid prototyping workloads [1], [2]. A part of the LLSC mission is to deliver new and innovative technologies and methods, enabling scientists and engineers to quickly ramp up the pace of their research. By leveraging supercomputing and big data storage assets, the LLSC has built the MIT SuperCloud, a coherent fusion of the four largest computing ecosystems: supercomputing, enterprise computing, big data, and traditional databases. The MIT SuperCloud has spurred the development of a number of cross-ecosystem innovations in high-performance databases [3], [4], database management [5], data protection [6], database federation [7], [8], data analytics [9] and system monitoring [10].

The MIT Center for Engaging Supercomputing's TX-E1 is a research system operated by the LLSC on behalf of the university partners of the Massachusetts Green High Performance Computing Center (MGHPCC) [11]. Use of TX-E1 is provided to faculty, staff, and students of the five founding universities and their external collaborators. We quickly acknowledged the potential workload of account creation and maintenance across this diverse user base and looked for ways to improve the situation. Federated authentication seemed to solve the most problems, allowing users to seamlessly utilize the authentication credential from their primary organization for access to our systems.

In this paper, we describe the tools used and process of enabling the MIT SuperCloud Portal for federated authentication with the InCommon Federation and the U.S. Government Public Key Infrastructure (PKI), the combination of which represents millions users with ties to academic institutions or the U.S. Government. We explore the software and methodology required to configure our system to accept credentials acquired by these two providers. Finally, we describe a number of security and user experience enhancements enjoyed by our users as a result of this deployment, the most notable of which is the ability to leverage robust, existing multi-factor authentication systems deployed by their primary institutions. While the initial focus of our efforts was various methods of web-based access to our supercomputing facilities, including our reverse-proxy forwarding services, we also describe a process for self-service registration and validation of secure shell (SSH) keys.

## II. MOTIVATION

Managing a large production HPC system with an extremely diversely affiliated user base spanning tens of unique organizations poses many obvious administrative challenges. The implementation of federated identity management allows us to offload a significant portion of the overhead involved in basic account maintenance. Routine tasks which consume a large fraction of administrator time, such as password resets, identity verification, and group membership assignment, can be eliminated almost entirely. In addition, we also gain significant new capabilities by tying each user's SuperCloud account with the account at their primary organization, including automatic account termination if a user ends their relationship with the organization we identify them with.

Multi-factor authentication is becoming commonplace for Internet-connected resources. However, there are many challenges with implementing it at scale [13]. Many of the federated authentication providers available already implement, and often even require, multi-factor authentication for all user sign-ins. While we have not yet chosen to implement a strict multi-factor authentication requirement on our own systems, and will likely always maintain a pool of exempt accounts for ad hoc classroom use, providing it as an option for the majority of account holders was highly desirable.

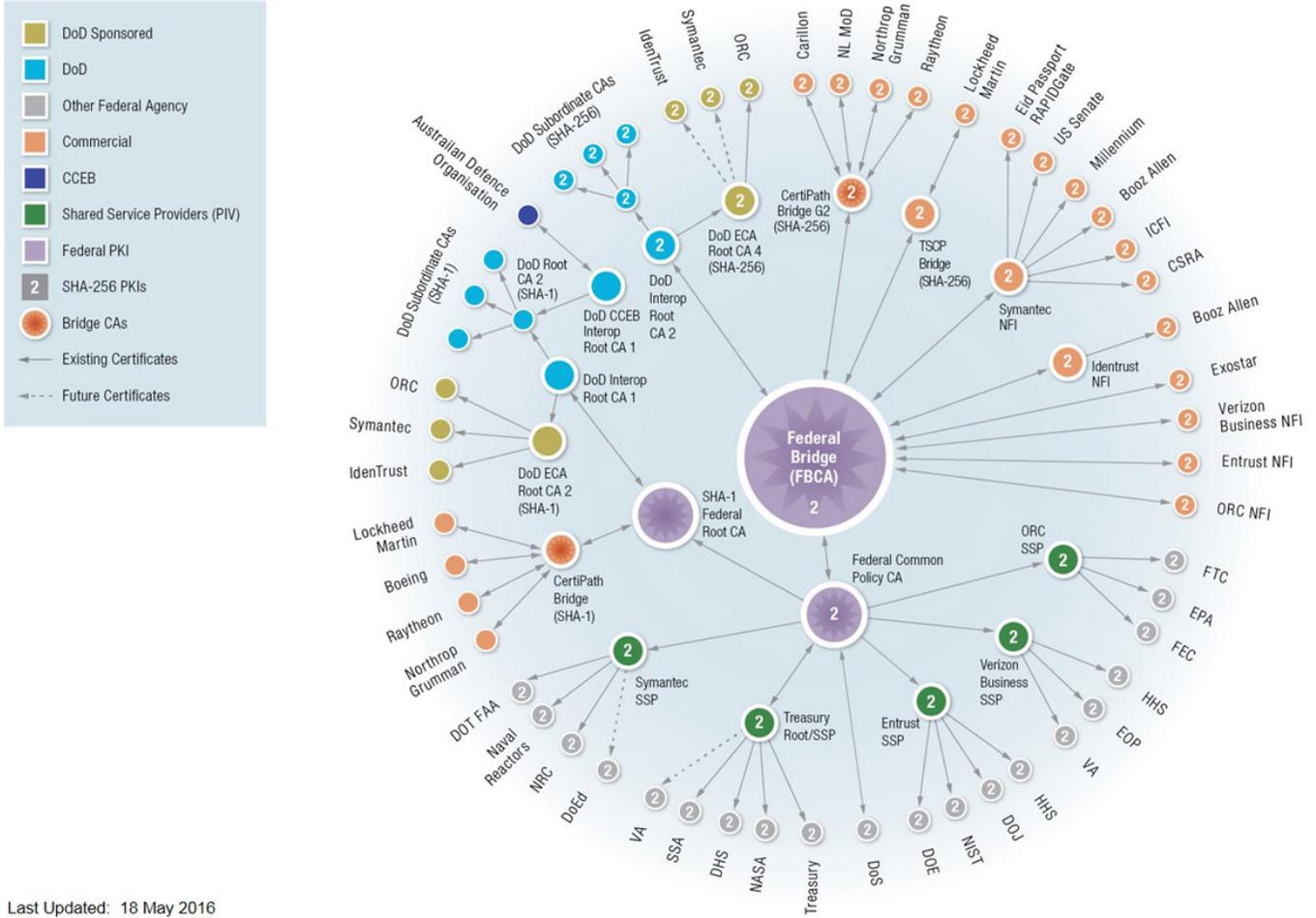

Fig. 1: The Federal PKI Interoperability Diagram taken from the Defense Information Systems Agency [12] illustrates the various interactions between the Department of Defense and external PKIs through the Federal Bridge.

We decided to integrate with two ecosystems of federated authentication and will discuss the motivations for each one separately.

*A. InCommon Federation*

The InCommon Federation is the U.S. education and research identity federation that provides a network of identity providers for cross-organizational single sign-on (SSO). More than 1000 organizations from higher education, research labs and sponsored partners participate [14].

The majority of accounts on our system are held by participants in the MGHPCC community. The most common primary organizational affiliation of the users on our system is MIT. Touchstone is the name of MIT's identity provider that interfaces with the InCommon Federation. All logins to MIT Touchstone require multi-factor authentication through Duo [15]. Many of the other InCommon participants also integrate with Duo, either providing it as an optionally enabled feature or a strict login requirement.

*B. U.S. Government PKI*

The U.S. Government maintains a large PKI web of trust between its various agencies and partners. Policies governing these relationships are issued by the Federal Public Key Infrastructure Policy Authority [16]. These policies allow for various levels of assurance in the issued PKI credential. The rigorousness of identity verification and the controls protecting private key information associated with the PKI certificate are the primary factors contributing to the level of assurance indicated for human subscribers. These levels of assurance policies are embedded within the issued certificate via the inclusion of X.660 [17] object identifiers (OIDs) called policy OIDs.

The U.S. Government has three distinct PKI programs belonging to this ecosystem: the Department of Defense (DoD) PKI, the Federal Common Policy (CP) PKI, and the Federal Bridge (FB) PKI. The DoD PKI supports the Common Access Card (CAC), which is a PKI smart card issued to

| user_ID | 1 |
| --- | --- |
| username | jdoe |
| email | jdoe@example.edu |
| active | ☑ |
| ts_principal | jdoe@example.edu |

Certificates:

| Remove? | cert_ID | CertSDN | NotAfter |
| --- | --- | --- | --- |
| ☐ | 1 | /C=US/O=U.S. Government/OU=DoD/OU=PKI/OU=CONTRACTOR/CN=Doe.John.1234567890 | 2019-11-21 17:00:00 |

New Certificate Mapping: [            ]

[Save Changes]

Fig. 2: MIT SuperCloud user edit page showing configuration options for mapping InCommon Federation users by their *eduPersonPrincipalName* or U.S. Government-approved PKI users by their Subject Distinguished Name.

all active-duty military, reservists, civilian employees, and many DoD contractors. Common Policy supports all U.S. Government agencies except the DoD and issues Personal Identity Verification (PIV) cards to all federal government employees and contractors in support of Homeland Security Presidential Directive 12 [18]. Finally, commercial entities may join the U.S. Government PKI ecosystem by entering into an agreement with the Federal Bridge. Agreements exist between these three programs so that each may accept credentials issued by either of the two others, depending on their indicated level of assurance. An agreement need not include all levels of assurance used by the PKIs entering into the agreement. These agreements are expressed in a machine-readable format through the issuance of cross-certificates, which include an extension that maps, or declares equivalency, between a specific policy OID of the issuer and subject PKI program. These agreements can express unilateral or bilateral trust, the distinction between which results in either one or two cross-certificates being issued. This cross-certificate is considered invalid for any policies not included in the policy map extension, reflecting the fact that it is not covered by the agreement between the organizations.

The result of these agreements is the large web of trust shown in Figure 1, as seen from the Department of Defense perspective. The DoD only accepts certificates meeting the Federal Bridge *mediumHardware* assurance policy (or stricter) which mandates identify verification using government issued photo IDs and a hardware key storage device for the private key. More than 5.4 million users hold active CAC or PIV cards issued by the U.S. Government [19] under the DoD or CP PKI. No estimates are available for the number of users holding PKI credentials from commercial entities participating in the FB PKI. It was clear to us that accepting PKI credentials from this ecosystem would immediately allow a very large user base to use their existing credentials to access our system. That the vast majority of these credentials are multi-factor PKI smart cards is an additional bonus.

### III. IMPLEMENTATION

We decided to use the MIT SuperCloud Portal as the cornerstone of our federated identity implementation. This technology grew out of a prototype built for Defense Research and Engineering (DR&E) [20]. This prototype was developed with a view toward facilitating the use of HPC resources from a segmented network with strict egress filtering; requirements included the ability to authenticate with the DoD CAC, to mount the central Lustre file system using common web ports via the WebDAV protocol [21], and to transparently submit HPC jobs using only these same standard ports using interactive HPC tools. In follow-on efforts, we enhanced this capability to add support for the Federal Common Policy and Federal Bridge PKIs, and the ability to extract the public key from a smart card for use with OpenSSH's PKCS#11 option added in 2010 (v5.4) or PuTTY-CAC [22], to provide multi-factor SSH authentication.

The SuperCloud portal technology is based on Apache httpd, and much of its functionality is implemented using a custom multi-processing module (MPM) that impersonates the web-authenticated user for all operating system calls. This initial design choice was necessary for file system access through the mod_dav module, which provides WebDAV services, to properly respect normal file system permissions. Conveniently, this design means that all other functions performed by the Apache web server, including regular web content serving, server-side script execution such as CGI or PHP scripts, and the watcher module described in our 2010 publication, also run with the system permissions of the web-authenticated user. This design ensures that users' actions through the web portal are limited to those they could perform if they opened an SSH session to the system.

In order to extend the portal to accept credentials supplied by a federated identity provider and map them to a local system user, we needed to develop two new features: a user interface component to prompt among a selection of multiple authentication providers and a mirror component on the server side to provide a pluggable interface to these providers and performs the actual authentication. Prior to the implementation of this project, the portal's authentication mechanism was built on HTTP basic access authentication backed by a Linux pluggable authentication modules (PAM). This is one of the simplest and most straightforward methods of authenticating a web browser to a web server, but comes with a number of downsides. With HTTP basic access authentication, the username and password entered by the user is cached by the browser and re-sent as part of the header in every HTTP request, and, upon receipt, the server performs validation of the user's identity. As a result, sessions are completely non-persistent; this is incompatible with the session persistence required by any token-based authentication scheme. To accommodate this session persistence requirement and improve architectural flexibility, we replaced our basic authentication framework with one based on web-form-based submission and implemented cookie-based session tracking while maintaining our use of Linux PAM for the initial user access verification. As we had previously used this mechanism for our DR&E prototype and other work that leveraged PKI smart card authentication, we were able to reuse much of our existing code in this implementation.

### A. InCommon Federation

The InCommon Federation is implemented using Secure Association and Markup Language (SAML) 2.0. SAML employs digitally signed XML messages passed between a service provider (SP) and an identity provider (IdP) to authenticate users. These messages typically also carry a bundle of attributes describing the user, including remote user name, email address, affiliation type, or other custom metadata. A service provider may trust one or more identity providers to authenticate its user base, generally resulting in a prompt for the user to select the appropriate one as part of the logon process. This SAML message exchange is depicted in Figure 3.

Entities in SAML are represented by metadata: an XML bundle containing the public cryptographic keys used to safeguard the integrity and authenticity of messages, the web addresses of SAML interfaces, entity names, and other basic information. The biggest challenge in federating SAML between many unrelated organizational entities is arranging discovery and exchange of this metadata. The InCommon Federation was created to solve this problem. InCommon centrally aggregates metadata about IdPs and SPs for all participating members into publications that it then signs with its own cryptographic key, providing updates following a set schedule. Upon joining the federation, a system administrator will statically configure the federation's public key, metadata download location, and update frequency. IdPs and SPs check these lists for changes according to this configured schedule and update their information cache once they have validated that the cryptographic signature matches the one statically configured for the federation.

Joining the InCommon Federation as an authorized service provider was straightforward; as MIT is already a member,

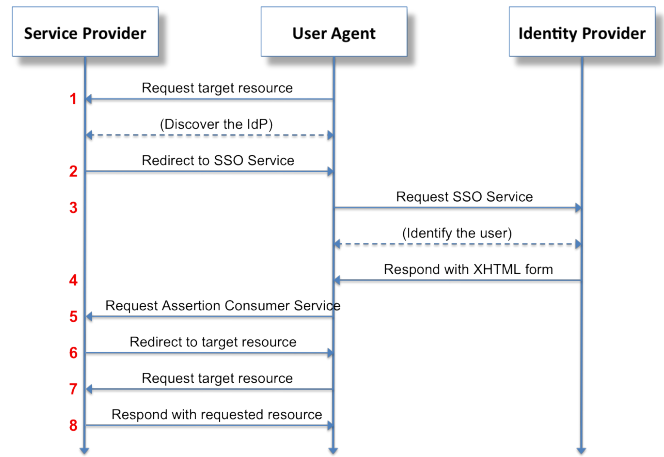

Fig. 3: SAML 2.0 Web Browser Single Sign-on: The service provider sends a digitally signed SAML Request to the identity provider's SSO service through the client web browser using an HTTP 301 redirect. The identity provider verifies that the service provider is authorized as part of its network, and the identity provider returns a digitally signed and/or encrypted SAML 2.0 response to the service provider Assertion Consumer Service using an HTTP POST through the client web browser. The signature and encryption certificates are validated by using the published metadata digitally signed by the InCommon Federation. SAML2 Browser SSO Redirect Post by Tom Scavo is licensed under CC BY-SA 3.0

becoming a subordinate entity within the greater web of trust simply involved coordinating with the MIT InCommon point of contact to have our SP metadata sent to InCommon for inclusion in their scheduled publications. In our case, this SP metadata consists of our entity name, access URL (the external Internet address of our web portal), our contact information, the user attributes we wish to be sent, and our X.509 certificate. Once published in the greater metadata feed, IdPs within member institutions could accept login requests from their users identifying our SP as the resource they wish to access. The IdP can then use our requested user attributes to identify the data that needs to be sent to our SP, the access URL to validate the browser POST back to our SP shown in Figure 3 step 4, and the X.509 certificate to validate the digital signature on the request and/or encrypt the response back to us.

Our first implementation task was to decide what SAML SP software to use on our system; as the SAML standard is widely adopted for federated identity management, there are myriad commercial and open-source products available to choose from. We chose the open-source SimpleSAMLphp project because our login processes were already written in PHP and the product is well supported by a team led by UNINETT, a state-owned company responsible for Norway's National Research and Education Network. We realized that it would be trivial to integrate this framework into our existing workflow and allow users to choose their preferred authentication method, verify their identity, and establish session persistence using cookies. Many of the other SAML authentication alternatives we investigated were deeply integrated with Apache and would deliver a great deal less implementation flexibility.

Configuring SimpleSAMLphp was pleasingly straightforward. To establish our service provider, we generated a key pair with an associated self-signed X.509 certificate and configured SimpleSAMLphp to set our server name, description, and contact information to embed in the SP metadata. Next, we configured the InCommon Federation metadata to automatically refresh on a set schedule using the metadata signing key and the download URL of the feed. We added a new button to our login page that imported the SimpleSAMLphp class and called the `requireAuth()` function followed by `getAttributes()`, at which point we're able to test if the *eduPersonPrincipalName* attribute is present and check its value against our local user account database seeking a match for a local system user name.

Next, we needed a method of populating the *eduPersonPrincipalName* in our local user database. For the MIT users who made up the bulk of our initial testers, this task was trivial: the value of this attribute was always the same as their email address. Unfortunately, we found that this was not true for many of our partner organizations, many of which either use a different identifier for this field, often a randomly generated string. We found that the fastest way to onboard these users is to have them attempt a login to our system, fail, and have a privileged user manually review an error log which provides sufficient information for us to identify the user and their associated *eduPersonPrincipalName*. Once validated, they then manually copy the value from the log to the ts_principal field of user's account, named after MIT's InCommon IdP service Touchstone, as shown in Figure 2.

We encountered an unexpected challenge early in the rollout of this system to the greater SuperCloud community: Many organizational identity providers are configured to disregard any list of requested attributes described earlier over concerns for their users' privacy. In these cases, the remote identity providers relay back only an informational message containing an empty set of attributes and the indication that an anonymous valid user from that organization had successfully logged in, with no hint to *which* user from that organization it was. The InCommon Federation acknowledged this problem and created the Research and Scholarship Entity Category (R&S) [23] in response, designating SPs that are "operated for the purpose of supporting research and scholarship interaction, collaboration or management." This category defines a more limited set of attributes that should be released by all participating IdPs for SPs with the R&S endorsement. Management of the R&S category has transitioned from InCommon to a Research and Education FEDerations group (REFEDS) working group to standardize the definition among the different research and education identity federations around the world. At the time of this writing, roughly one-third of the IdPs in the InCommon Federation support the R&S category attribute release.

Applying for inclusion in the R&S category was as straightforward as our initial enrollment in the federation and effectively solved all remaining issues for the roughly one-third of the IdPs in the InCommon Federation that support the R&S category attribute release. For the remaining IdPs that do not automatically release needed attributes upon request or honor the R&S category, we have needed to individually contact the IdP point of contact at each specific entity to request permission, typically upon first acceptance of a user account request from a member of that organization. This has thus far been a simple, albeit manual process, never taking more than a few days. Every organization we have worked with has enabled the release globally and not on a per-user basis; subsequent user sign-ups from the same organization are able to proceed without manual intervention.

*B. PKI*

PKI client certificate authentication, including the optional use of a smart card for storage of the private key, is built into all modern web browsers. A configurable option is available to servers whereby, as part of the TLS handshake, they may request that clients present an X.509 certificate as a means of client identification in addition to the mandatory process of the server sending the client a certificate to identify itself. We chose to use a separate virtual host with its own unique subdomain name for this purpose because of the inconsistency with which different web servers and browsers handle per-directory TLS renegotiation. Once the SuperCloud Portal was converted to use cookie-based session tracking, we simply needed to reintegrate existing code for PKI-X based certificate authentication from our previously mentioned work.

The PKI standards have supported cross-certificates and policy mapping since 1999 [24]. The potential for cross-certificates creates multiple valid paths between an end-user certificate and self-signed trust anchors. Discovering, building, and validating these paths, a process known as path discovery and path validation, can be very difficult. Implicit in the web-of-trust nature of PKI federations, bidirectional cross-certificates are to be expected, creating the possibility of loops in the certificate chains. Revoked and reissued cross-certificates further complicate the possibilities. Many applications, including Apache httpd, do not implement a path discovery and validation algorithm capable of correctly processing complex PKI federations.

Previous efforts undertaken in this area used statically configured certificate path discovery, and we reapplied that strategy here. While not ideal, as the PKI standard envisioned path discovery being a dynamic activity, we found the volatility of the PKI certificate paths in organizations we worked with to be very low and quite manageable to address with a static configuration. A static certificate path discovery configuration provides benefits in speed and simplicity of implementation, with the downside of this approach being that manual intervention is required for the first user of a new organization that has a valid cross-certificate in the U.S. Government PKI ecosystem that we have not previously configured.

Our approach to path discovery was further motivated by its potential simplification of our second challenge: path valida-tion. Certificate path validation involves checking the digital signature and certificate revocation status of all certificates involved in an authentication request, beginning with the end-user's certificate and proceeding until we reach a configured trust anchor. There are two possible methods of querying the revocation status of a certificate: a certificate revocation list (CRL), which provides a complete list of all invalidated certificates issued by that certificate authority, and the Online Certificate Status Protocol (OCSP), which allows for interactive queries regarding the status of a specific certificate. The uniform resource identifier for either or both of these methods can be listed in the issued certificate.

Our current implementation only supports the older, but more ubiquitous, CRL method. By choosing to use CRLs for path validation along with statically configured paths, we are able to download all the necessary revocation information in advance into a local cache, eliminating any background download tasks that could delay the user's interactive login. As some of the Department of Defense CRLs, in particular, have grown excessively large in the past because of the size of the DoD PKI user base, these download delays could potentially result in an unpleasantly sluggish user login experience during peak times if they are not mitigated with a local cache.

The final step is the ability to map valid certificates presented by existing users to their associated accounts on our system. We map these certificates to accounts using the Subject Distinguished Name (SDN) field embedded within the certificate. Similarly to the way we handled new InCommon registrations, here we also found that the fastest way for users to inform us of their certificates' SDN is to have them attempt to log in, fail, and have that failure deliver sufficient information for us to identify users. Privileged users are provided with tools to quickly associate this metadata provided by failed logins with existing system users, as shown in Figure 2. While other methods exist, such as looking up a user in the DoD PKI 411 database or instructing the user how to navigate the user interface to discover this information, in practice these methods have always proven to be less time efficient.

*C. Secure Shell Access*

We have seen user demand for Secure Shell (SSH) access to our systems decline over time. While still required for the majority of users, it is no longer the absolute requirement it once was. Through our web-based portal, users have the ability to launch Jupyter Notebooks, Accumulo databases, and virtual machines [25], and in the context of a running Jupyter Notebook or Lab session, users can access a web-based terminal window that behaves very similarly to a SSH session. We also provide the ability to securely forward arbitrary web applications running on compute resources out through the portal, while simultaneously preventing inadvertent or unauthorized access with a user-based firewall on the internal HPC network [26]. Users can access the web services they launch through our forwarding proxy by using either a subdirectory URL rewriting scheme, if supported by their web service, or a sub-domain forwarding for the growing list of web services that do not. This combination of access methods fully satisfies the needs of a growing fraction of our user base.

Our policy for SSH access has always been to require the use of public key authentication. Prior to this federated authentication effort, users needed to email our support team a copy of their public key for addition to the system to establish their initial access. Using the impersonation ability of the portal, it was trivial to write a self-service web page providing the ability for users to edit their own SSH authorized_keys file once they authenticated with their federated identity. This web portal also allows us to sanity check the submitted keys for appropriate cryptographic strength and known weaknesses in generation [27], [28].

IV. CONCLUSIONS

This paper details an effort to integrate with two large authentication ecosystems that provide external identity verification for millions of users. We describe our experiences transitioning to federated authentication as our primary authenticator for the MIT SuperCloud TX-E1 system's web-based user portal. Through our portal's ability to impersonate the authenticated user for system calls, we are able to provide web-based remote access to users' files via the WebDAV protocol and the ability for them to interactively launch arbitrary HPC jobs. The portal's forwarding reverse proxy technology allows for seamless access to users' custom web applications running on the cluster, while enforcing security though integration with our user-based firewall. While we notice an increasing fraction of our user base no longer requests SSH access to the cluster, users are also provided the option to register and manage SSH public keys associated with their account in a self-service manner. The combination of these capabilities results in a high-productivity environment to users of all HPC experience levels.

By having users authenticate with their existing credentials from their primary organization, we have eliminated most of the administrative overhead involved in user account management from our team and removed users' burden of maintaining separate credentials. We have also enhanced the security of our system by taking advantage of the existing multi-factor authentication systems deployed by some of these organizations. Ninety-three percent of our active user base, defined as having submitted an HPC job in the last six months and excluding temporary classroom accounts, have been enrolled for federated authentication, and all new user accounts are created with federated authentication enabled. We have received only positive feedback from our users.